\documentclass[aps,prb,twocolumn,amsmath,amssymb,superscriptaddress,showpacs]{revtex4}

\usepackage{graphicx}

\usepackage{bm}

\def\be{\begin{equation}}

\def\ee{\end{equation}}

\def\bea{\begin{eqnarray}}

\def\eea{\end{eqnarray}}

\newcommand{\matr}[4]{{\left(\begin{array}{cc} #1&#2\\#3&#4\\\end{array}\right)}}

\newcommand{\vect}[2]{{\left(\begin{array}{c} #1\\#2\\\end{array}\right)}}

\renewcommand{\vec}{\mathbf}

\renewcommand{\vr}{\vec{r}}

\newcommand{\vsigma}{\mbox{\boldmath $\sigma$}}

\newcommand{\vA}{\vec{A}}

\newcommand{\vB}{\vec{B}}

\newcommand{\vnabla}{\mbox{\boldmath $\nabla$}}

\newcommand{\vp}{\vec{p}}

\newcommand{\ket}[1]{|#1\rangle}

\newcommand{\bra}[1]{\langle#1|}

\newcommand{\sgn}{\mathrm{sgn}}

\newcommand{\eps}{\varepsilon}

\newcommand{\Tr}{\mathrm{Tr}}

\newcommand{\vF}{v_{\rm F}}

\begin{document}

\title{Local sublattice symmetry breaking for graphene with a centro-symmetric deformation}

\author{M. Schneider}
\affiliation{Dahlem Center for Complex Quantum Systems and Institut f\"ur Theoretische Physik, Freie Universit\"at Berlin, Arnimallee 14, 14195 Berlin, Germany}
\author{D. Faria}
\affiliation{Instituto de F\'{\i}sica, Universidade Federal Fluminense, Niter\'oi,
Av.\ Litor\^anea sn 24210-340, RJ-Brazil}
\author{S. {Viola Kusminskiy}}
\affiliation{Dahlem Center for Complex Quantum Systems and Institut f\"ur Theoretische Physik, Freie Universit\"at Berlin, Arnimallee 14, 14195 Berlin, Germany}
\author{N. Sandler}
\email[N. Sandler: ]{sandler@ohio.edu}
\affiliation{Department of Physics and Astronomy, Nanoscale and Quantum Phenomena Institute, Ohio University, Athens, Ohio 45701-2979, USA}
\date{\today}


\begin{abstract}
We calculate the local density of states (LDOS) for an infinite graphene sheet with a single centro-symmetric out-of-plane deformation, in order to investigate measurable strain signatures on graphene. We focus on the regime of small deformations and show that the strain-induced pseudomagnetic field induces an imbalance of the LDOS between the two triangular graphene sublattices in the region of the deformation. Real space imaging reveals a characteristic six-fold symmetry pattern where the sublattice symmetry is broken within each fold, consistent with experimental and tight-binding observations. The open geometry we study allows us to make use of the usual continuum model of graphene and to obtain results independent of boundary conditions. We provide an analytic perturbative expression for the contrast between the LDOS of each sub-lattice, showing a scaling law as a function of the amplitude and width of the deformation. We confirm our results by a numerically exact iterative scattering matrix method. 
\end{abstract}

\maketitle

\noindent{\it Introduction.---} Graphene under strain has been largely discussed in the literature and explored for different geometries, with particular features providing alternative routes to confine and control its charge carriers~\cite{Pereira,Vozmediano,Peeters}.
Significant development in the theoretical description of strained graphene elucidated how its electronic properties are modified on strained surfaces. At the microscopic level, a general deformation is described by modifications in the atomic positions which reflects in the Hamiltonian as local changes in the hopping parameter~\cite{CastroNeto,Papa}. In the continuum model these changes appear as an effective gauge field, and electrons with momentum around the Dirac valleys move in the deformed region as in the presence of a pseudomagnetic field~\cite{Ando}. Strain also produces a deformation potential, i.e.,  a scalar field similar to a local chemical potential that can affect electron dynamics~\cite{Ando}.  Very recently, measurements in high-quality graphene samples on particular substrates suggested a strong connection between random fluctuations in strain and transport properties~\cite{Morpurgo}.

The use of strain effects to engineer graphene electronic properties has also been explored in several experiments in the last years~\cite{Levy, Klimov, Mashoff, Tomori, JiongLu, Georgiou}. As one of the most relevant findings, Levy et al. were able to show the presence of pseudo Landau levels generated by giant pseudomagnetic fields induced by homogeneous strain in graphene nanobubbles~\cite{Levy}. This experimental confirmation that strain can have striking effects on the electronic properties of graphene has been followed by other experiments that explore the effect of different geometries as a path to control graphene electromechanically~\cite{Tomori, JiongLu, Klimov, Mashoff}. A generic deformation of a graphene sheet can cause {\it inhomogeneous} strain, which results in an effective non-uniform pseudomagnetic field and provides an experimental test-bed to explore the interplay between highly tunable magnetic fields and Dirac fermions. For example, a scanning tunneling microscope (STM) tip has been 
used not only to probe samples, but also to continuously deform graphene nanomembranes, demonstrating electronic confinement due to non-uniform pseudomagnetic fields~\cite{Klimov}.  For a similar experimental setup, Mashoff et al. obtained atomically resolved STM images of stable and lifted regions of graphene~\cite{Mashoff}. Whereas a hexagonal arrangement of the carbon atoms was found at unstrained regions, as expected for monolayer graphene~\cite{Mashoff}, within the strained area a triangular pattern of bright spots was observed, signaling a symmetry breaking between A and B sublattices in some regions. At the time the authors speculated the effect to be caused by an instability in which the different sublattices acquire a zigzag configuration with respect to the substrate. However, a local sublattice rearrangement requires energies that are prohibitive within the regime of STM imaging making such scenario rather unlikely.  Atomistic tight-binding models\cite{Sloan,Barraza,Ramon} have predicted the 
development of such asymmetry but a continuum, symmetry-based description remains missing. Similar patterns were observed in earlier works but remained unexplained also\cite{Xu}. These results indicate an incomplete understanding of the fundamental electronic properties of graphene samples where local manipulation produce effective inhomogeneous gauge and scalar fields.

In this work, we approach this problem  by investigating the electronic properties of a graphene sheet in the presence of an axially symmetric out-of-plane
deformation. The strain produced by such distortion is represented by a pseudomagnetic field with trigonal symmetry and embodies a good approximation to standard experimental configurations, while still allowing for analytical treatment. We  use a scattering formalism based on the continuum description of graphene to address the question of confinement of electrons due to this deformation. In particular we calculate the LDOS and show that a noticeable imbalance in the distribution of charge density between the two graphene nonequivalent sublattices appears even for small deformations, providing a possible explanation for the experimentally observed sublattice asymmetry. We perform exact numerical calculations and show that these results are well described within an analytical perturbative approach for small deformations. We analyze the dependence of the maximum LDOS contrast between sublattices on energy and strain strength, providing a scaling dependence with the parameters of the deformation. 
Finally, we also show that the effective scalar field introduced by strain minimally modifies the predicted sublattice asymmetry. 

\noindent{\it Model.---} The electronic properties of undeformed graphene are, for low energies and large system sizes, governed by two copies of a two-dimensional (2D) Dirac Hamiltonian $H_0=\vF \vp\cdot \vsigma$
where $\vF\approx 10^6 m/s$ is the velocity of graphene electrons, $\vp$ the electronic momentum around the K (K') point, and $\vsigma=(\sigma_x,\sigma_y)$ are Pauli matrices reflecting the pseudospin degree of freedom associated with the sublattice structure of the honeycomb lattice~\cite{Castro}. The  strain is produced by a mechanical deformation modeled with a height-profile $h(\vr)$ that is centro-symmetric  and is written generically as $h(\vr)=A\, h_0(r/b)$,
where $h_0$ contains the radial profile, and the parameters $A$ and $b$ describe amplitude and effective radius of the deformation. In the following, to illustrate our results, we consider the case of a Gaussian bump with height profile $h_0(x)=e^{-x^2}$. Note however that our results below are qualitatively valid for a generic profile $h_0$ with axial symmetry.

The effect of such deformation on the electronic properties in the continuum limit are described within the theory of elasticity. For an out-of-plane deformation the strain tensor of elasticity~\cite{Landau} is derived from the height profile $h$ according to $\epsilon_{ij}=\frac{1}{2} \partial_i h \partial_j h$, which in polar coordinates ($r,\theta$) reads
\begin{equation}
\label{eq:elasticity}
 \epsilon=\alpha f(r/b)\matr{\cos^2\theta}{\sin\theta\cos\theta}{\sin\theta\cos\theta}{\sin^2\theta},
\end{equation}
where $\alpha=A^2/b^2$ sets the strength of the strain, while its spatial distribution is contained in the function $f(x)=\frac{1}{2}\left[h_0'(x)\right]^2$. For the Gaussian profile, one has $f(x)=2x^2 e^{-2x^2}$. 

In the presence of the deformation, electrons experience the strain as a gauge field
\begin{equation}
\label{eq:vectorpot}
 \vA(\vr)=-\frac{g_v}{e\vF} \alpha f(r/b) \vect{\cos 2\theta}{-\sin 2\theta},
\end{equation}
where we chose the zigzag direction to lie along the $x$-axis~\cite{Vozmediano}. The coupling constant is $g_v=\beta \hbar \vF/2 a\approx 7 eV$\cite{CastroNeto}, being $\beta=|\partial \log t/\partial \log a|\approx 3$, and $t$ and $a$ the hopping parameter and the lattice constant of graphene.

For the radial symmetric deformation, the associated pseudomagnetic field $\vB=\vnabla\times \vA$ shares the trigonal symmetry of the graphene lattice,
\begin{equation}
 B_z(\vr)=\frac{\hbar}{e} \left(\frac{-\beta}{2ab}\right) \alpha b_0(r/b) \sin 3\theta
\end{equation}
where the spatial profile is given by the function $b_0(x)=\frac{2f(x)}{x}-f'(x)$. For the Gaussian-shaped deformation, one has $b_0(x)=8x^3 e^{-2 x^2}$.

\begin{figure}[t]
\includegraphics[width=2.7in]{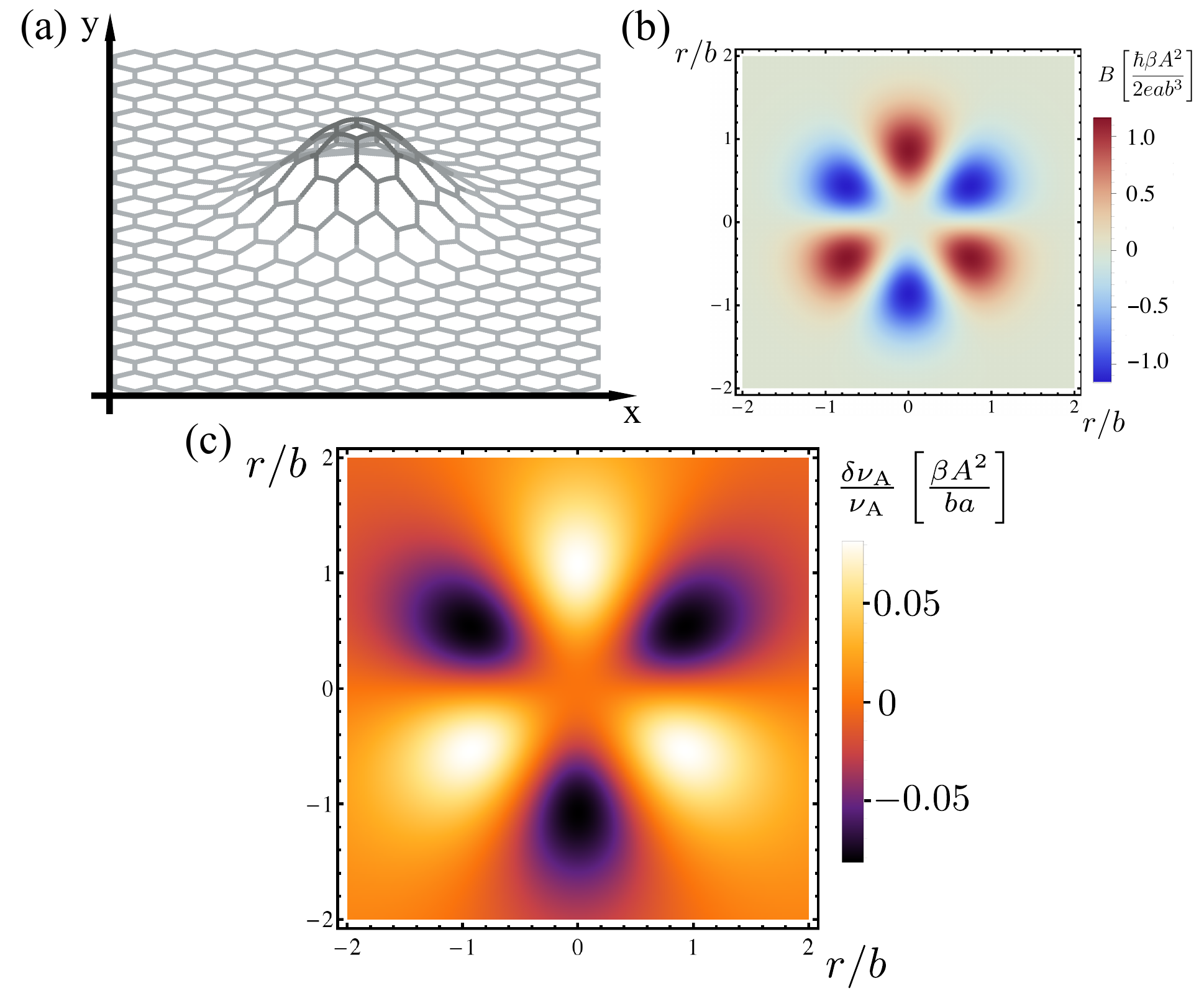}
\caption{(Color online) Schematic view of the graphene lattice with a magnified out-of-plane deformation. (b) Pseudomagnetic field created by a deformation with a Gaussian height profile as in (a). (c)  Spatial profile of the LDOS for sublattice A in the presence of a bump, see Eq.\ \eqref{eq:ldos}. Bright (dark) spots indicate an increase (decrease) of LDOS compared to the undeformed graphene sheet. For sublattice B, the effect is exactly opposite.}
\label{fig:bfield}
\end{figure}

In addition to the gauge field, the electrons are also exposed to a scalar potential proportional to the trace of the strain tensor.  In our model it is represented by $V(\vr)=-g_s \alpha f(r/b)$ with $g_s=3 eV$ as a typical value\cite{Sloan}. 
The low-energy electronic properties in the presence of the deformation are hence described by 
\begin{equation}
 H=\vF \left[\vp+e \vA(\vr)\right]\cdot \vsigma + V(\vr)
\end{equation}
In this article we consider a bump that is smooth on the scale of the lattice constant, such that a coupling between the valleys can be neglected~\cite{Sloan,Ramon, Blanter, Bubaloo}. Moreover, we consider an infinite graphene sheet containing a single deformation, hence our results are independent of finite size effects and boundary conditions\cite{Bubaloo,Blanter,Barraza,Ramon,Sloan}. 

\noindent{\it Perturbation theory.---} In this section we present analytic results for the change in the LDOS produced by the scattering of electrons off the deformation obtained with a perturbative approach in real space. We consider therefore small deformations that allow for an expansion in the parameter $\alpha$. 

From now on we set $\hbar=\vF=1$, and work around the K valley. The effect of the K' valley is discussed at the end of this section. We split the Hamiltonian in the kinetic part $H_0$  and the perturbation ${\cal V}$, 
\begin{equation}
 \label{eq:calV}
 {\cal V}(\vr)= e \vA(\vr) \cdot \vsigma=\matr{0}{A_-(\vr)}{A_+(\vr)}{0},
\end{equation}
where we defined $A_{\pm}(\vr)=e(A_x(\vr)\pm i A_y(\vr))=\left(\frac{-\beta}{2a}\right) \alpha f(r/b) e^{\mp2 i \theta}$. We neglect the scalar potential in this part, we will include its effect in the next section.

Let us start with the states of the Dirac equation in the absence of the deformation. Here, we take circular waves as a set of basis states,
\begin{equation}
 \ket{\Phi^{(0)}_m(\vr)}=\sqrt{\frac{\eps}{4\pi}} e^{i m \theta} \vect{e^{-i\theta/2} J_{|m-1/2|}(\eps r)}{i\,\sgn(m) e^{i\theta/2} J_{|m+1/2|}(\eps r)},
 \label{eq:psi0}
\end{equation}
where $\eps$ denotes the energy of the Dirac fermions (which we assume to be positive, for simplicity), and $m$ is a half-integer index labeling the states according to their angular momentum. $J_{n}(x)$ denotes the Bessel function of $n$-th order. We chose a normalization such that 
\begin{equation}
 \label{eq:norm}
 \int d\vr \langle\Phi^{(0)}_m(\eps;\vr) \ket{\Phi^{(0)}_n(\eps';\vr)} =\delta_{nm} \delta(\eps-\eps').
\end{equation}
Our goal is to find the scattering state $\ket{\Psi_m(\vr)}$, that replaces $ \ket{\Phi^{(0)}_m(\vr)}$ when the bump is present. This is determined by the Lippmann-Schwinger equation
\begin{equation}
\label{eq:Lippmann}
 \ket{\Psi_m(\vr)}=\ket{\Phi^{(0)}_m(\vr)}+\int d\vr' G(\vr,\vr') {\cal V}(\vr') \ket{\Psi_m(\vr')}
\end{equation}
which contains the Green's function of graphene,
\begin{equation}
\label{eq:Green}
 G(\vr,\vr')=-i\pi \sum_m \begin{cases}
                             \ket{\Phi^{(0)}_m(\vr)} \bra{\Phi^{(-)}_m(\vr')}, & r<r'\\
                              \ket{\Phi^{(+)}_m(\vr)} \bra{\Phi^{(0)}_m(\vr')}, & r>r',
                           \end{cases}
\end{equation}
where we defined
\begin{equation}
 \ket{\Phi^{(\pm)}_m(\vr)}=\sqrt{\frac{\eps}{4\pi}} e^{i m \theta} \vect{e^{-i\theta/2} H^{(\pm)}_{|m-1/2|}(\eps r)}{i\,\sgn(m) e^{i\theta/2} H^{(\pm)}_{|m+1/2|}(\eps r)}.
\end{equation}
$H^{(\pm)}_{\mu}(x)=J_{\mu}(x)\pm i Y_{\mu}(x)$ are Hankel functions of first and second kind. A derivation of Eq.~\eqref{eq:Green} is given in the Supplementary Material\cite{Supp}.

For small deformations, we solve the Lippmann-Schwinger equation in the Born approximation, and replace the scattering state $\ket{\Psi_m(\vr)}$ on the right-hand side of the equation by the unperturbed state $ \ket{\Phi^{(0)}_m(\vr)}$. Note that our perturbative approach is valid for  $g_{s,v} \alpha \ll \eps$. The explicit form of the resulting scattering states is shown in the Supplementary Material\cite{Supp}. The trigonal symmetry of the pseudomagnetic field underlies the coupling between angular momentum states differing by $3$.

The LDOS is obtained by calculating $\nu(\eps,\vr)=\sum_m \langle \Psi_m(\vr) \ket{\Psi_m(\vr)}$. The new states are properly normalized to leading order in $\alpha$, as the linear in $\alpha$ correction is orthogonal to the unperturbed state. 
Since we are interested in the different sublattice occupations, we further introduce the sublattice-resolved LDOS
\begin{equation}
 \label{eq:LDOS_Sub}
 \nu_{\rm A/B}(\eps,\vr)=\sum_m \bra{\Psi_m(\vr)} {\cal P}_{\rm A/B} \ket{\Psi_m(\vr)}
\end{equation}
where ${\cal P}_{\rm A/B}$ are projectors on the respective sublattice ${\rm A/B}$.
For undeformed graphene, evaluating Eq.~\eqref{eq:LDOS_Sub} with the free states $ \ket{\Phi^{(0)}_m(\vr)}$ produces the well-known value of 
\begin{equation}
 \nu_{\rm A,B}^{(0)}(\eps,\vr)=\frac{\eps}{4\pi} \sum_m \left[J_{|m-1/2|}(\eps r)\right]^2=\frac{\eps}{4\pi},
 \label{eq:ELDOSAB}
\end{equation}
for the LDOS per sublattice. 

We now want to discuss the effect of the deformation on the LDOS. Specifically we address the limit $\eps b\ll 1$, which is the relevant case for experiments with a radius of a few lattice constants. In this case, one may simplify the results by using the asymptotic expressions of the Bessel and Hankel functions for small arguments $r\lesssim b$~\cite{Supp}. Upon retaining only the leading contribution for small energies, one finds for the corrections $\delta \nu_{\rm A,B}=\nu_{\rm A,B}-\nu^{(0)}_{\rm A,B}$ to leading order in $\alpha$
\begin{equation}
 \label{eq:ldos}
 \frac{\delta \nu_{\rm A}(\eps,r)}{\nu_{\rm A}(\eps,r)}=-\frac{\delta \nu_{\rm B}(\eps,r)}{\nu_{\rm B}(\eps,r)}= - \frac{\beta A^2}{ b a}   \sin 3\theta \, g(r/b)
\end{equation}
with the function
\begin{align}
\label{eq:gx}
g(x)=\frac{1}{x^3}\int_0^{x} dy\, y^3\, f(y)
\end{align}
To leading order in $\alpha$, one can replace $\nu^{(0)}_{\rm A,B}$ by $\nu_{\rm A,B}$ in the denominator of Eq.~\eqref{eq:ldos}. Notice that the relative LDOS correction has no dependence on energy. Thus, {\it the deformation changes the local occupation in the different sublattices in opposite directions}, and their spatial distribution shares the symmetry of the underlying pseudomagnetic field with a radial distribution governed by the function $g(x)$. Specifically for a Gaussian height profile, one finds 
\begin{align}
\label{eq:gx_gauss}
 g(x)=\frac{1}{4 x^3} \left[1-e^{-2 x^2}(1+ 2x^2+2x^4)\right].
\end{align}
The spatial distribution of the change in LDOS for sublattice A according to Eq.~\eqref{eq:ldos} is shown in Fig.~\ref{fig:bfield}~(c) for a Gaussian deformation of typical dimensions.

\begin{figure}[t]
\includegraphics[width=2.7in]{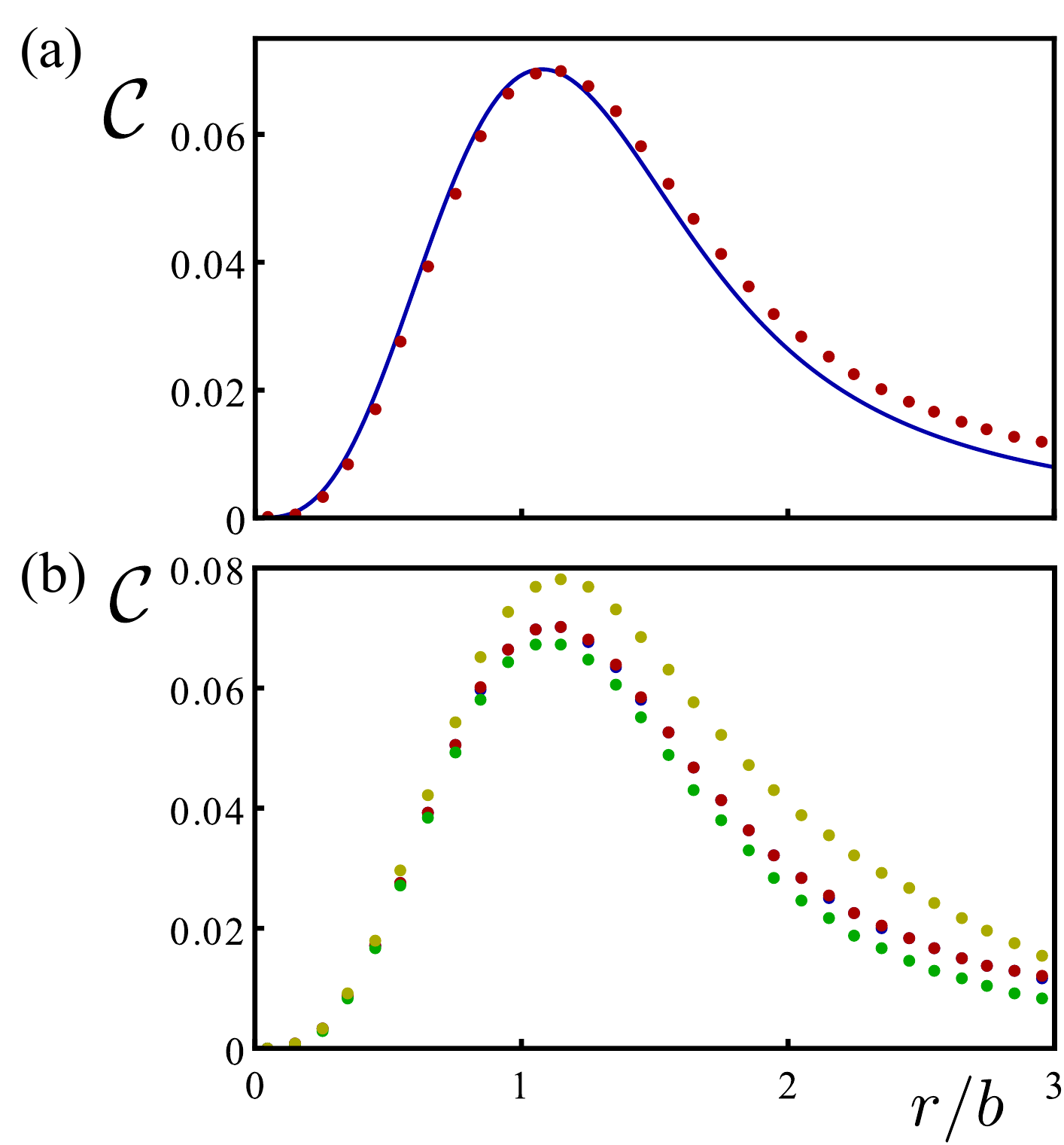}
\caption{(Color online) LDOS contrast ${\cal C}$ as a function of distance from the bump's center for fixed angle $\theta=\pi/2$ ($A=0.1 nm, b=0.5 nm$).  (a) Comparison  of ${\cal C}$ between perturbative (solid line, blue), and exact numerical approaches (red points) for $\eps=0.5 eV$. (b)  Different data sets obtained numerically for $\eps=0.5 eV, g_s=0$ (blue), $\eps=0.5 eV, g_s= 3 eV$ (red), $\eps=0.1 eV, g_s=3 eV$ (green), and $\eps=0.9 eV, g_s=3 eV$ (yellow).}
\label{fig:pertnum}
\end{figure}

\begin{figure}[t]
\includegraphics[width=2.7in]{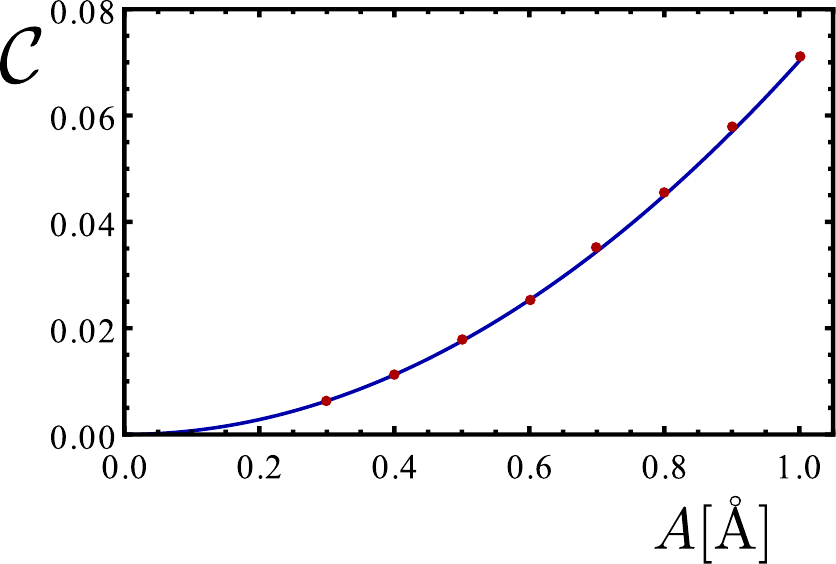}
\caption{(Color online) Scaling of contrast ${\cal C}$ as a function of amplitude for fixed Gaussian width $b=0.5 nm$. Curves obtained with perturbative (solid line, blue), and exact numerical methods (points, red) for $\eps=0.5 eV$  are compared.}
\label{fig:ContrastVsA}
\end{figure}

A quantity of experimental relevance is the LDOS contrast between sublattices, defined as
\begin{equation}\label{eq:contrast}
 {\cal C}=2\frac{|\nu_{\rm A}-\nu_{\rm B}|}{\nu_{\rm A}+\nu_{\rm B}}
\end{equation}
which is plotted as a function of the radial distance in Fig.~\ref{fig:pertnum}~(a). Note that, for a fixed width $b$ of the deformation, Eq~\eqref{eq:ldos} indicates that the contrast ${\cal C}$ scales quadratically with the amplitude of a centro-symmetric deformation. This scaling is shown for a Gaussian deformation in Fig.~\ref{fig:ContrastVsA} and compared with the exact numerical results presented in the next section.

To conclude this section let's discuss the role of the two valleys K and K' in these results.  As mentioned above, the deformation is smooth enough that does not couple the valleys and their contributions add directly. To see that these are identical, note that the Dirac Hamiltonian takes the same form in both valleys when the spinors are written in the valley symmetric representation\cite{Beenakker}: $(\psi_{A}, \psi_{B})$ around valley K, and $(\psi_{B},-\psi_{A})$ around valley K'. Note that the components referring to $A$ and $B$ sublattices are interchanged between different valleys. On the other hand, the pseudomagnetic field enters the Dirac equation with opposite sign for each valley (in contrast to a real magnetic field that has the same sign in both valleys). These two effects ensure that their contributions to the sublattice occupancy contrast are identical.

\noindent{\it Numerics.---} In this section we discuss briefly our exact numerical approach for the continuum model, which is not restricted to the case of small amplitude deformations. The results obtained confirm our findings described in the previous section in the corresponding regimes. We use a slight modification of the method introduced in Ref.~\onlinecite{Martin}, which allows for the calculation of the scattering matrix $S$ for an arbitrarily-shaped scalar potential. The extension to include a vector potential is straightforward. To calculate the LDOS integrated over a certain (arbitrarily chosen) volume $V$ per sublattice, we include a fictitious additional scattering potential
\begin{equation}
 {\cal V}_{\eps}(\vr)=\matr{\eps_{\rm A}}{0}{0}{\eps_{\rm B}} \times \begin{cases}
                                                                                   1, & \vr \in V\\
										   0, & \mbox{else} 
                                                                                  \end{cases}
\end{equation}
Such potential locally changes the electron's energy  in $V$ by a magnitude $\eps_{\rm A/B}$ in the different sublattices ${\rm A/B}$. The LDOS per sublattice integrated over $V$ is then found from the scattering matrix $S$ via
\begin{equation}
 \nu_{j}(V)=\frac{1}{2\pi i} \Tr \left[S^{\dagger} \frac{\partial S}{\partial \eps_{j}} \right]_{\eps_{A/B}=0}
\end{equation}
where $j={\rm A,B}$ specifies the sublattice. Fig.~\ref{fig:pertnum}~(a) shows a comparison between analytical and numerical results for realistic parameters. Note that in the region of small amplitudes the contrast obtained with the expression from perturbation theory follows closely the exact solution given by the numerical approach. Numerical calculations in the presence of the scalar potential $V(\vr)$ induced by the deformation were carried out for different values of the phenomenological parameter $g_s$. Our results, as shown in Fig.~\ref{fig:pertnum}~(b), confirm that its effect on the contrast is negligible (note that, to leading order in $\alpha$, $V(\vr)$ affects the occupation of both sublattices in the same way, thus not affecting the contrast~\eqref{eq:contrast}). Furthermore, the energy independent value for the contrast predicted with the analytical approach (as long as the requirements $\eps\gg \alpha g_{s,v}$ and $\eps b\ll 1$ are met) is also verified in the numerical results as shown in 
Fig.~\ref{fig:pertnum}~(b).

\noindent{\it Conclusions.---} We have shown that a centro-symmetric deformation, with its local breaking of the lattice translational symmetry, produces a {\it local sublattice symmetry breaking} on the electronic properties of a graphene sheet, and a consequent LDOS contrast between sublattices. Analytic expressions within the Born approximation predict the intensity of the LDOS contrast to be determined by the amplitude of the deformation and to be energy independent for the range of validity of the approximation. Exact numerics carried out with scattering matrix methods confirm the validity of these results, for experimentally realistic parameters. While our numerical approach allows us in principle to treat any size of deformations, we concentrated here on the study of small deformations and showed that there is a measurable LDOS contrast between sublattices even in the absence of Landau levels~\cite{Castro}. The crossover to this regime 
will be published elsewhere. Our findings here provide an alternative interpretation for recent experimental observations on STM graphene nanomembranes~\cite{Alex} and provide a quantitative way to guide the use of strain in the design of electronic properties of graphene flakes.

{\it Acknowledgments.---} We gratefully acknowledge support by CNPq, and DAAD (D.F.), DFG SPP 1459 and the A. v H. Foundation (M.S., S.V.K.), and NSF No. DMR-1108285 (D.F., N.S.) as well as discussions with R. Carrillo-Bastos, A. Georgi, M. Morgenstern and P. Nemes-Incze. 

\appendix

\section{Greens functions}\label{app:GF}

In this appendix, we show a derivation of the Green function for graphene given in Eq.\ 9 in the main text. The defining equation reads
\begin{equation}
 (\eps+i \vsigma \cdot \vnabla) G(\vr,\vr')=\delta(\vr-\vr').
\end{equation}
For the derivation, it is customary to introduce the auxiliary Green function $G_s(\vr,\vr')$ via
\begin{equation}
 \label{eq:Ggraphene-Gscalar}
 G(\vr,\vr')=(\eps-i\vsigma \cdot \vnabla) G_s(\vr,\vr')
\end{equation}
The advantage of the Green function $G_s$ is that it satisfies a scalar equation ({\it i.e.} the pseudospin degree of freedom has been eliminated).
\begin{equation}
 (\eps^2+\vnabla^2) G_s(\vr,\vr') = \delta(\vr-\vr')
\end{equation}
Moreover, $G_s$ is the solution to the Helmholtz equation in two dimensions. 
We would like to express the Green function in polar coordinates, thus we express the Laplacian as
\begin{equation}
 \Delta=\partial_r^2+\frac{1}{r}\partial_r+\frac{1}{r^2}\partial_{\theta}^2,
\end{equation}
and we write for the $\delta$-function
\begin{equation}
 \delta(\vec{r}-\vec{r}')=\frac{1}{r}\delta(r-r')\sum_{\mu}\frac{1}{2\pi}e^{i\mu(\theta-\theta')},
\end{equation}
where the summation is running over integer $\mu$. We then expand $G_s$ as
\begin{equation}
 G_{s}(r,\theta;r',\theta')=\sum_{\mu}\frac{1}{2\pi} e^{i\mu(\theta-\theta')} g_{\mu}(r,r')
\end{equation}
where $g_{\mu}(r,r')$ is determined by
\begin{equation}
 \left(\eps^2+\partial_r^2+\frac{1}{r}\partial_r-\frac{\mu^2}{r^2}\right)g_{\mu}(r,r')=\frac{1}{r} \delta(r-r').
\end{equation}
For $r\neq r'$, $g_{\mu}$ is thus satisfying a Bessel differential equation. We therefore write
\begin{equation}
 g_{\mu}(r,r')= a_{\mu} J_{|\mu|}(\eps r_<) H^{(+)}_{|\mu|}(\eps r_>)
\end{equation}
with $r_>=\mathrm{max}(r,r')$, $r_<=\mathrm{min}(r,r')$. In this way, we respect the conditions, that 
\begin{itemize}
 \item $g_{\mu}$ is regular at $r \rightarrow 0$,
 \item $g_{\mu}$ is behaving as an outgoing wave ($\propto \frac{e^{i\eps r}}{\sqrt{r}}$) for $r\rightarrow \infty$ (this is the proper boundary condition for the retarded Green function $\eps\rightarrow \eps+i0$),
 \item $g_{\mu}$ is continuous at $r=r'$.
\end{itemize}
The remaining coefficient $a_{\mu}$ is determined by the jump condition for the derivative of $g_{\mu}$,
\begin{equation}
 \left[\partial_r g_{\mu} (r,r')\right]^{r=r'+0}_{r=r'-0}=\frac{1}{r'},
\end{equation}
which yields $a_{\mu}=\frac{\pi}{2 i}$. Thus, the result for the scalar Green function reads
\begin{equation}
 G_{s}(r,\theta;r',\theta')=\frac{1}{4i}\sum_{\mu} e^{i\mu(\theta-\theta')} J_{|\mu|}(\eps r_<) H^{(+)}_{|\mu|}(\eps r_>).
\end{equation}

The Green function for graphene is then found in a straightforward albeit lengthy calculation from Eq.\ \eqref{eq:Ggraphene-Gscalar}.

\section{Scattering states}\label{app:scattstates}

The scattering states in the Born approximation are obtained by inserting Eq.\ 5, Eq.\ 6 and Eq.\ 9 into the following equation 
\begin{equation}
\label{eq:Born}
 \ket{\Psi_m(\vr)}=\ket{\Phi^{(0)}_m(\vr)}+\int d\vr' G(\vr,\vr') {\cal V}(\vr') \ket{\Phi^{(0)}_m(\vr')}.
\end{equation}
After insertion of the corresponding expressions, one finds
\begin{widetext}
\begin{eqnarray}
 \ket{\Psi_m(\vr)}=&&\ket{\Phi^{(0)}_m(\vr)} \nonumber \\
&-&i\pi\sum_{n} \ket{\Phi^{(+)}_n(\vr)}\int_{r'<r} r'dr' \int d\theta' \bra{\Phi^{(0)}_n(\vr)} {\cal V}(\vr') \ket{\Phi^{(0)}_m(\vr')}  \nonumber \\ 
&-&i\pi\sum_{n} \ket{\Phi^{(0)}_n(\vr)}\int_{r'>r} r'dr' \int d\theta' \bra{\Phi^{(-)}_n(\vr)} {\cal V}(\vr') \ket{\Phi^{(0)}_m(\vr')}.
\end{eqnarray}
The angular integration is nonzero for $n=m+3$ and $n=m-3$.  Calculating the matrix elements, one obtains the scattering states given by
\begin{align}
 \label{eq:psiscatt}
 \ket{\Psi_m(\vr)}=&\ket{\phi_m^{(0)}(\vr)} +\pi\frac{\beta}{4a} \eps\, \alpha\left(a_m(r) \ket{\phi_{m+3}^{(+)}(\vr)}+b_m(r) \ket{\phi_{m-3}^{(+)}(\vr)}+c_m(r) \ket{\phi_{m+3}^{(0)}(\vr)} +d_m(r) \ket{\phi_{m-3}^{(0)}(\vr)}  \right)
\end{align}
with
\begin{align}
a_m(r)=&-\sgn(m) \int_0^{r} dr' r' f(r'/b) J_{|m+5/2|}(\eps r')    J_{|m+1/2|}(\eps r') \nonumber\\
b_m(r)=&        \sgn(m-3)            \int_0^{r} dr' r' f(r'/b) J_{|m-5/2|}(\eps r')    J_{|m-1/2|}(\eps r') \nonumber\\
c_m(r)=&       - \sgn(m) \int_r^{\infty} dr' r' f(r'/b) H^{(+)}_{|m+5/2|}(\eps r')     J_{|m+1/2|}(\eps r') \nonumber\\
d_m(r)=&                \sgn(m-3)     \int_r^{\infty} dr' r' f(r'/b) H^{(+)}_{|m-5/2|}(\eps r')    J_{|m-1/2|}(\eps r'). \nonumber\\
\end{align}

The leading order correction to the local density of states per sublattice up to first order on $\alpha$, when the deformation is present,  is given by
\begin{eqnarray}
\frac{\delta \nu_A}{\nu_A}\left(\eps,\vr\right)=&-&\pi \frac{\beta}{4a}\eps\alpha e^{3i\theta}\sum_{m} J_{|m-1/2|}(\eps r) H^{(+)}_{|m+5/2|}(\eps r)a_m(r) \nonumber \\ 
&+&\pi \frac{\beta}{4a}\eps\alpha e^{-3i\theta}\sum_{m} J_{|m-1/2|}(\eps r) H^{(+)}_{|m-7/2|}(\eps r)b_m(r) \nonumber \\ 
&-&\pi \frac{\beta}{4a}\eps\alpha e^{3i\theta}\sum_{m} J_{|m-1/2|}(\eps r) J_{|m+5/2|}(\eps r)c_m(r) \nonumber \\
&+&\pi \frac{\beta}{4a}\eps\alpha e^{-3i\theta}\sum_{m} J_{|m-1/2|}(\eps r) J_{|m-7/2|}(\eps r)b_m(r) + c.c,
\end{eqnarray}
\renewcommand{\arraystretch}{1.7}
with $\nu_A=\frac{\eps}{4\pi}$ for clean graphene. Using the asymptotic forms of the Bessel functions for small arguments
\begin{equation}
J_n(x)\approx \frac{1}{2^n n!}x^n,
\end{equation}
\begin{equation}
Y_n(x)\approx \Biggl\{\begin{array}{c}\frac{2}{\pi}\text{ln}\left(\frac{e^\gamma}{2}x\right), \,\,\,\,\,\,\,\,\,     n=0 \\
-\frac{2^n(n-1)!}{\pi x^n}, \,\,\,\,\,  n \geq1 \end{array}
\end{equation}
one finds the dominant contribution to $\delta \nu_A$ from the second line, for m=1/2 (+c.c.):
\begin{equation}
 \frac{\delta \nu_{\rm A}}{\nu_{\rm A}}= - \frac{\beta b}{a}\alpha   \sin 3\theta \, g(r/b).
\end{equation}
The LDOS correction for sublattice B is $\delta \nu_B/\nu_B= -\delta \nu_A/\nu_A$. 
\end{widetext}

\end{document}